\documentclass[letterpaper, 10 pt, conference]{ieeeconf}
\IEEEoverridecommandlockouts
\usepackage{cite}
\usepackage{amsmath,amssymb,amsfonts}
\usepackage{algorithmic}
\usepackage{graphicx}
\usepackage{textcomp}
\usepackage{xcolor}
\usepackage{lipsum}
\usepackage{cuted}
\usepackage{array}
\usepackage{algorithm}
\usepackage{dsfont}
\usepackage[caption=false,font=footnotesize]{subfig}

\title{\LARGE \bf Age of View: A New Metric for Evaluating Heterogeneous Information Fusion in Vehicular Cyber-Physical Systems}

\author{Xincao Xu$^{1}$, Kai Liu$^{1}$, Qisen Zhang$^{1}$, Hao Jiang$^{1}$, Ke Xiao$^{2}$ and Jiangtao Luo$^{3}$
\thanks{$^{1}$Xincao Xu, Kai Liu, Qisen Zhang, and Hao Jiang are with the College of Computer Science, Chongqing University, Chongqing, China ({\tt\small e-mail: {\{near, liukai0807, qszhang, artanis\}@cqu.edu.cn}}) }
\thanks{$^{2}$Ke Xiao is with the College of Computer and Information Science, Chongqing Normal University, Chongqing, China
      ({\tt\small e-mail: {xiaoke@cqnu.edu.cn}})}
\thanks{$^{3}$Jiangtao Luo is with the School of Communication and Information Engineering, Chongqing University of Posts and Telecommunications, Chongqing, China
      ({\tt\small e-mail: {luojt@cqupt.edu.cn}})}
}

\begin{document}
\maketitle
\thispagestyle{empty}
\pagestyle{empty}

\begin{abstract}
Heterogeneous information fusion is one of the most critical issues for realizing vehicular cyber-physical systems (VCPSs).
This work makes the first attempt at quantitatively measuring the quality of heterogeneous information fusion in VCPS by designing a new metric called Age of View (AoV).
Specifically, we derive a sensing model based on a multi-class M/G/1 priority queue and a transmission model based on Shannon theory.
On this basis, we formally define AoV by modeling the timeliness, completeness, and consistency of the heterogeneous information fusion in VCPS and formulate the problem aiming to minimize the system's average AoV.
Further, we propose a new solution called Multi-agent Difference-Reward-based deep reinforcement learning with a Greedy Bandwidth Allocation (MDR-GBA) to solve the problem. 
In particular, each vehicle acts as an independent agent and decides the sensing frequencies and uploading priorities of heterogeneous information.
Meanwhile, the roadside unit (RSU) decides the Vehicle-to-Infrastructure (V2I) bandwidth allocation for each vehicle based on a greedy scheme.
Finally, we build the simulation model and compare the performance of the proposed solution with state-of-the-art algorithms. 
The experimental results conclusively demonstrate the significance of the new metric and the superiority of the proposed solution.
\end{abstract}

\begin{keywords}
Vehicular cyber-physical system, Heterogeneous information fusion, Deep reinforcement learning	
\end{keywords}

\section{Introduction}

Benefitting from current advances in sensing technologies and vehicular communications, the development of vehicular cyber-physical systems (VCPSs) \cite{jia2015survey} has received great attention in both industries and academia, which aims to enable the next generation of intelligent transportation systems (ITSs) by bridging the gap between the physical vehicular environment and the cyber system view.
With VCPS, heterogeneous information such as the status of traffic lights, vehicle locations, point cloud data, and traffic surveillance videos, could be synergistically sensed and uploaded, so as to construct logical views of the physical environment and facilitate the implementation of emerging ITS applications.

Great efforts have been devoted to enhancing system performance in vehicular networks, including data dissemination, information caching, task offloading, etc.
Liu et al. considered the cooperative data dissemination problem in a vehicular end-edge-cloud architecture and proposed a clique searching-based scheduling scheme to enable collaborative data encoding and transmission \cite{liu2020fog}.
Singh et al. proposed an intent-based network control framework, in which a  neural network is used to train the flow table and enable intelligent data dissemination \cite{singh2020intent}.
Dai et al. proposed a blockchain-empowered distributed content caching framework, where vehicles perform content caching and base stations maintain the blockchain \cite{dai2020deep}.
Xiao et al. developed a binary particle swarm optimization based coding scheduling to exploit synergistic effects of network coding and vehicular caching \cite{xiao2021cooperative}. 
Shang et al. studied energy-efficient computation offloading in VEC and developed a deep-learning-based algorithm to optimize transmission power and computation resources \cite{shang2021deep}.
Liao et al. presented an intent-aware task offloading strategy, which enables vehicles to learn the long-term strategy under information uncertainty with multi-dimensions intent awareness \cite{liao2020learning}.

A number of studies have focused on developing VCPS, including the prediction, scheduling, and control of system status.
Zhang et al. proposed a hybrid velocity-profile prediction method, which integrates traffic flow state with individual driving behaviors \cite{zhang2019cyber}.
Zhang et al. predicted vehicle status based on lane-level localization and acceleration prediction \cite{zhang2020data}.
A lane-change behavioral prediction model and an acceleration prediction model are derived based on historical driving data.
Lian et al. presented a scheduling method for vehicle path planning based on an established map model \cite{lian2020cyber}.
Liu et al. proposed a scheduling algorithm for temporal data dissemination in VCPS \cite{liu2014temporal}.
Liu et al. presented a temporal data scheduling method considering the dynamic snapshot consistency requirement in VCPS \cite{liu2014scheduling}.
Xu et al. proposed a vehicle collision warning scheme based on trajectory calibration by considering V2I communication delay and packet loss \cite{xu2020vehicular}.
Lv et al. presented an adaptive algorithm to control vehicle acceleration under three typical driving styles with different protocol selections \cite{lv2018driving}.

Distinguishing from the above efforts, this work is dedicated to designing a new metric for evaluating the quality of heterogeneous information fusion in VCPS and proposing a new solution to maximize system performance.
The critical issues to be addressed are summarized as follows.
First, the information uploaded by vehicles is time-varying, which has to be sensed and updated in time.
So, the synergistic impact of sensing frequency, queuing delay, and transmission delay has to be considered to keep the information fresh.
Second, due to the high mobility of vehicles, limited Vehicle-to-Infrastructure (V2I) communication range, and unreliable wireless transmissions, how to enhance the completeness of the information at the roadside unit (RSU) is another critical concern.
Third, since the information is intrinsically heterogeneous in terms of distribution and updating frequency, it is important yet challenging to construct a uniform representation of different pieces of information.

The main contributions are outlined as follows. 
\begin{itemize}
\item We propose a new metric called Age of View (AoV) to quantitatively measure the quality of heterogeneous information fusion in VCPS. First, the multi-class M/G/1 priority queue is adopted to model the information queuing sensed by vehicles. Then, the Shannon theory is applied to model the packet delay and loss via V2I communication. On this basis, we formally describe the uniform representation of heterogeneous information. Finally, the AoV is defined by modeling the timeliness, completeness, and consistency of information fusion. 
\item We propose a new solution called Multi-agent Difference-Reward-based deep reinforcement learning with Greedy Bandwidth Allocation (MDR-GBA), which enables distributed sensing at vehicles and centralized bandwidth allocation at the RSU. Specifically, vehicles act as independent agents and decide the sensing frequencies and uploading priorities of heterogeneous information. Then, a greedy bandwidth allocation scheme is designed to allocate V2I bandwidth by considering vehicle mobility and ITS application requirements.
\item We build the simulation model and compare the performance of the proposed algorithm with competitive deep reinforcement learning based solutions, including Centralized Deep Deterministic Policy Gradient (C-DDPG)\cite{lillicrap2015continuous}, Multi-agent Actor-Critic (MAC)\cite{lowe2017multi}, and a variation of MAC called MAC-GBA. Realistic vehicle trajectories extracted from Didi GAIA Initiative \cite{didi} are adopted for performance evaluation. The simulation results conclusively demonstrate the significance of the new metric AoV and the superiority of the proposed solution. In particular, MDR-GBA outperforms C-DDPG, MAC, and MAC-GBA by around 16.43\%, 16.71\%, and 4.91\%, respectively, in terms of maximizing the cumulative system reward.
\end{itemize}

The rest of this paper is organized as follows.
Section II designs the new metric AoV.
Section III proposes the new solution MDR-GBA.
Section IV evaluates the system performance.
Finally, Section V concludes this work.

\section{Age of View Formulation}
\subsection{Notations}
The set of discrete time slots of the system is denoted by $\mathcal{T}=\left\{\tau_{1}, \tau_{2}, \cdots, \tau_{t}, \cdots, \tau_{|\mathcal{T}|}\right\}$, where $|\mathcal{T}|$ is the number of time slots, and the time slot length is denoted by $\varepsilon$.
The set of vehicles is denoted by $S=\left\{s_{1}, s_{2}, \cdots, s_{i}, \cdots, s_{|S|}\right\}$, where $|S|$ is the number of vehicles.
The heterogeneous information is divided into $J$ categories and each category has K data items, which is denoted by $D^{J \times K}$, where $d_{j k}$ is the $k$-th information of the $j$-th category.
The size of $d_{j k}$ is denoted by $\left|d_{j k}\right|$.
The RSU is denoted by $e=\left\langle l_{e}, r_{e}, b_{e}\right\rangle$, where $l_{e}$, $r_{e}$, and $b_{e}$ denote the location, communication range, and bandwidth of the RSU $e$, respectively.
The distance between vehicle $s_i$ and RSU $e$ is denoted by a function of time $\operatorname{dis}_{i e}^{t} \triangleq \operatorname{distance}\left(l_{i}^{t}, l_{e}\right)$, where $\operatorname{distance}\left(\cdot,\cdot\right)$ is the Euclidean distance. 

\subsection{System Model}
\begin{figure}
\centering
\includegraphics[width=0.99\columnwidth]{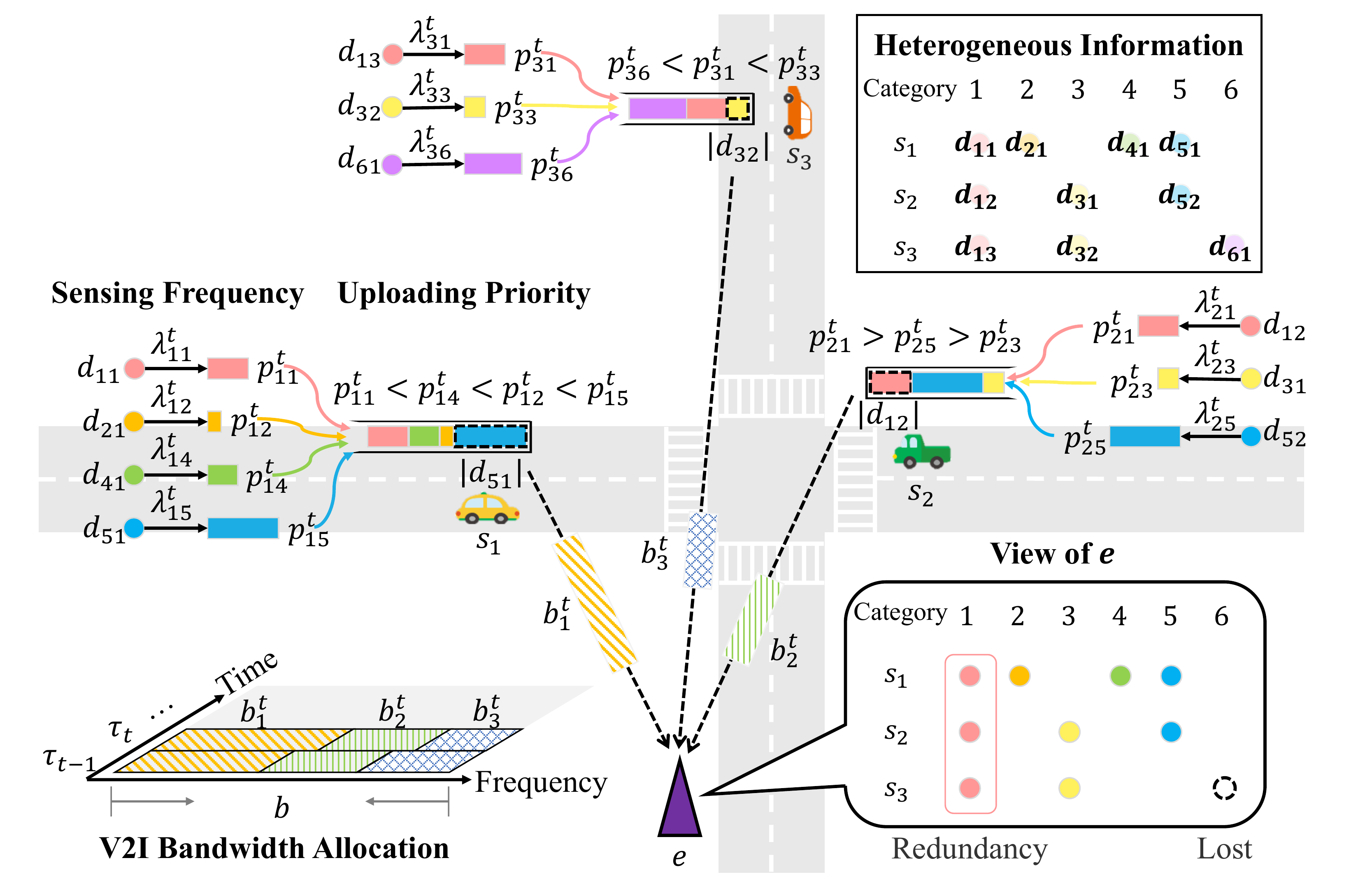}
  \caption{System model}
  \label{fig_1_system_model}
\end{figure}

As shown in Fig. \ref{fig_1_system_model}, the queuing time of each category of heterogeneous information sensed by vehicles is modeled by multi-class M/G/1 priority queue \cite{moltafet2020age}.
The information sensed by vehicle $s_i$ is denoted by $D_i=\left\{D_{i 1}, D_{i 2}, \cdots, D_{i j}, \cdots \right\}$, where $D_{i j}$ is the set of the $j$-th category information sensed by vehicle $s_i$.
The arrival rate of $D_{i j}$ in vehicle $s_i$ at time $\tau_{t}$ is determined by the sensing frequency and denoted by $\lambda_{i j}^{t}$.
Due to the limited sensing ability, we have $0<\lambda_{i j}^{\min } \leq \lambda_{i j}^{t} \leq \lambda_{i j}^{\max }, \forall D_{i j} \subseteq D_{i}, \forall s_{i} \in S, \forall \tau_{t} \in \mathcal{T}$, where $\lambda_{i j}^{\min }$ and $\lambda_{i j}^{\max }$ are the minimum and maximum of sensing frequency for the $j$-th category information in vehicle $s_i$, respectively.
The inter-arrival time ${int}_{i j}$ of $D_{i j}$ is the interval of arrival time between adjacent data items of the $j$-th category in vehicle $s_i$, which is computed by ${int}_{i j}=\frac{1}{\lambda_{i j}^{t}}$.

The uploading priority of $D_{i j}$ at time $\tau_{t}$ is denoted by $p_{i j}^{t} \in(0,1), \forall D_{i j} \subseteq D_{i}, \forall s_{i} \in S, \forall \tau_{t} \in \mathcal{T}$.
The information with higher priority will be uploaded earlier to the RSU.
The transmission time of data items in $D_{i j}$ from vehicle $s_i$ to the RSU $e$ follows a General distribution with mean $\mathds{E}\left[\operatorname{ser}_{i j}\right]$ and finite second moment $\mathds{E}\left[\operatorname{ser}_{i j}^{2}\right]$.
The uploading workload for vehicle $s_i$ can be represented by $\rho_{i}^{t}=\sum_{\forall D_{i j} \subseteq D_{i}} \lambda_{i j}^{t} \cdot \mathds{E}\left[\operatorname{ser}_{i j}\right]$.
To guarantee the existence of the queue steady-state, it requires $\rho_{i}^{t} < 1$.
The overall workload of $D_{i j^{*}}$, which has higher uploading priority than $D_{i j}$, is denoted by $\rho_{i j}^{t}=\sum_{\forall D_{i j^{*}} \subseteq D_{i}} \mathds{1}\{p_{i j^{*}}^{t} \geq p_{i j}^{t}\} \cdot \lambda_{i j^{*}}^{t} \cdot \mathds{E}\left[\operatorname{ser}_{i j^{*}}\right]$, where $\mathds{1}\{p_{i j^{*}}^{t} \geq p_{i j}^{t}\}$ is an indictor function, $\mathds{1}\{p_{i j^{*}}^{t} \geq p_{i j}^{t}\} = 1$, if $p_{i j^{*}}^{t} \geq p_{i j}^{t}$, otherwise, $\mathds{1}\{p_{i j^{*}}^{t} \geq p_{i j}^{t}\} = 0$.
Then, the average queuing time ${wai}_{i j}$ of $D_{i j}$ can be computed by:
\begin{equation}
\text{\footnotesize${wai}_{i j}=\frac{1}{1-\rho_{i j}^{t}+\lambda_{i j}^{t} \mathds{E}\left[\operatorname{ser}_{i j}\right]}\left[\mathds{E}\left[\operatorname{ser}_{i j}\right]+\frac{\mu_{i j}^{t}}{2\left(1-\rho_{i j}^{t}\right)}\right] -\mathds{E}\left[\operatorname{ser}_{i j}\right]$}
\end{equation}
where $\mu_{i j}^{t}=\sum_{\forall D_{i j^{*}} \subseteq D_{i}} \mathds{1}\{p_{i j^{*}}^{t} \geq p_{i j}^{t}\} \cdot \lambda_{i j^{*}}^{t} \cdot \mathds{E}\left[\operatorname{ser}_{i j^{*}}^{2}\right]$.

Then, we model the packet transmission time and the successful transmission condition via V2I communications based on Shannon theory.
The set of vehicles within the radio coverage of RSU $e$ at time $\tau_{t}$ is denoted by $S_{e}^{t}=\left\{s_{i} \mid \operatorname{dis}_{i e}^{t} \leq r_{e}, \forall s_{i} \in S\right\}$, $S_{e}^{t} \subseteq S$.
The bandwidth of vehicle $s_i$ allocated by RSU $e$ at time $\tau_t$ is denoted by $b_{i}^{t}$], and we have $0 \leq b_{i}^{t} \leq b_{e}, \forall s_{i} \in S_{e}^{t}, \forall \tau_{t} \in \mathcal{T}$, and $\sum_{\forall s_{i} \in S_{e}^{t}} b_{i}^{t} \leq b_{e}$, where $b_{e}$ is the bandwidth of RSU $e$.
The Signal to Noise Ratio (SNR) \cite{sadek2009distributed} of V2I communications between vehicle $s_i$ and RSU $e$ at time $\tau_{t}$ is computed by:
\begin{equation}
\label{equ_SNR}
\operatorname{SNR}_{i}^{t}=\frac{1}{N_{0}} \cdot\left|h_{i e}\right|^{2} \cdot \psi \cdot {\operatorname{dis}_{i e}^{t}}^{-\varphi} \cdot \pi
\end{equation}
where $N_{0}$ is the additive white Gaussian noise, $h_{i e}$ is the channel fading gain, $\psi$ is a constant that depends on the antennas design, $\varphi$ is the path loss exponent, and $\pi$ is the transmission power.
Due to the high mobility of vehicles, the SNR of signal received at RSU $e$ from vehicle $s_i$ may fall below an SNR threshold, which is called SNR wall \cite{tandra2008snr} and calculated by $\mathrm{SNR}_{\text {wall }}=\frac{\sigma^{2}-1}{\sigma}$, where $\sigma$ is a function of noise uncertainly $N_{0}^{*}$ measured by dB, $\sigma=10^{N_{0}^{*} / 10}$.

According to the Shannon theory, the achievable transmission rate $\delta_{i}^{t}$ of V2I communications between vehicle $s_i$ and RSU $e$ at time $\tau_{t}$ is calculated by $\delta_{i}^{t}=b_{i}^{t} \cdot \log _{2}\left(1+\operatorname{SNR}_{i}^{t}\right)$.
The transmission time ${tra}_{i j}$ of $d_{j k} \in D_{i j}$ from vehicle $s_i$ to RSU $e_k$ is calculated by ${tra}_{i j}=\frac{\left|d_{j k}\right|}{\delta_{i}^{t}}$.
A successful transmission refers to the event that the received SNR is above the SNR wall during the packet transmission.
Therefore, the successful transmission condition of $d_{j k}$ is represented by:
\begin{equation}
\label{Equ_pro_packet_loss}
\mathds{P}_{i j}^{t}=\left\{\begin{array}{l}
1, \forall \tau_{t^{*}} \in\left[\tau_{t}, \tau_{t} + {tra}_{i j}\right], \operatorname{SNR}_{i}^{t^{*}}>\mathrm{SNR}_{\text {wall }}\\
0, \exists \tau_{t^{*}} \in\left[\tau_{t}, \tau_{t} + {tra}_{i j}\right], \operatorname{SNR}_{i}^{t^{*}} \leq \mathrm{SNR}_{\text {wall }}
\end{array}\right.
\end{equation}

\subsection{The Metric of AoV}
The set of views is denoted by $V=\left\{v_{1}, v_{2}, \cdots, v_{g}, \cdots, v_{|V|}\right\}$.
The information requirement for view construction is denoted by $v_{g} \in\{0,1\}^{|S| \times {J}}$, where $v_{i j}^{t} = 1$ indicates that the construction of view $v_{g}$ requires data item in $D_{i j}$ sensed by vehicle $v_{i}$ at time $\tau_{t}$, otherwise, $v_{i j}^{t} = 0$.
The set of required views by the RSU $e$ at time $\tau_{t}$ is denoted by $V_{e}^{t} \subseteq V$. 
Then, we define the three characteristics of heterogeneous information including timeliness, completeness, and consistency as follows.
\newtheorem{Definition}{Definition}
\begin{Definition}[\textit{Timeliness}]
	The timeliness $\theta_{g} \in [0,+\infty)$ of view $v_{g}$ is defined as the sum of inter-arrival time, queuing time, and transmission time of each heterogeneous information received at the RSU.
	\begin{equation}
	\label{equ_timeliness}
	\theta_{g}=\sum_{\forall v_{i j}^{t} \in v_{g}} v_{i j}^{t} \cdot \mathds{P}_{i j}^{t} \cdot \left({int}_{i j}+{wai}_{i j}+{tra}_{i j}\right)
	\end{equation}
\end{Definition}
\begin{Definition}[\textit{Completeness}]
	The completeness $\chi_{g} \in [0,1]$ of view $v_g$ is defined as the ratio between received and required heterogeneous information of the view.
	\begin{equation}
	\chi_{g}=\frac{\sum_{\forall v_{i j}^{t} \in v_{g}} v_{i j}^{t} \cdot \mathds{P}_{i j}^{t}}{\sum_{\forall v_{i j}^{t} \in v_{g}} v_{i j}^{t}}
	\end{equation}
\end{Definition}
\begin{Definition}[\textit{Consistency}]
	 The consistency $\xi_{g} \in [0,+\infty)$ of view $v_g$ is defined as the quadratic sum of the difference between information received time and average information received time.
\begin{equation}
\xi_{g}=\sum_{\forall v_{i j}^{t} \in v_{g}}\left|{wai}_{i j}+{tra}_{i j}-\frac{\sum_{\forall v_{i j}^{t} \in v_{g}} {wai}_{i j}+{tra}_{i j}}{\sum_{\forall v_{i j}^{t} \in v_{g}} v_{i j}^{t} \cdot \mathds{P}_{i j}^{t}}\right|^{2}
\end{equation}
\end{Definition}

Finally, we give the formal definition of AoV, which synthesizes the three critical characteristics to measure the quality of a view.
\begin{Definition}[\textit{Age of View, AoV}]
	It is defined as a weighted average of normalized timeliness, completeness, and consistency of view $v_g$.
	\begin{equation}
	\operatorname{AoV}_{g}=\mathrm{W_{1}}  \hat{\theta_{g}}+\mathrm{W_{2}}(1-\chi_{g})+\mathrm{W_{3}} \hat{\xi_{g}}
	\end{equation}
\end{Definition}
\noindent where $\hat{\theta_{g}}$ and $\hat{\xi_{g}}$ are the normalized timeliness and consistency using min-max scaler formula; $\mathrm{W_{1}}, \mathrm{W_{2}}, \mathrm{W_{3}}$ are weighting factors for the normalized timeliness, completeness, and normalized consistency of view, respectively, and we have $\mathrm{W_{1}}+\mathrm{W_{2}}+\mathrm{W_{3}}=1$.

Given a solution $x = (\bf\Lambda, \mathbf{P}, \mathbf{B} )$, where $\mathbf{\Lambda}$ denotes the determined sensing frequencies and is represented by $\mathbf{\Lambda}=\left\{\lambda_{i j}^{t} \mid \forall D_{i j} \in D_{i}, \forall s_{i} \in S, \forall \tau_{t} \in \mathcal{T}\right\}$; $\mathbf{P}$ denotes the determined uploading priorities and is represented by $\mathbf{P}=\left\{p_{i j}^{t} \mid \forall D_{i j} \in D_{i}, \forall s_{i} \in S, \forall \tau_{t} \in \mathcal{T} \right\}$, and $\mathbf{B}$ denotes the determined V2I bandwidth allocation and is represented by $\mathbf{B}=\left\{b_{i}^{t} \mid \forall s_{i} \in S_{e}^{t}, \forall \tau_{t} \in \mathcal{T}\right\}$, the objective is to minimize the average AoV in the scheduling period $\mathcal{T}$, which is represented by:
\begin{equation}
	\begin{aligned}
		&\min _{\mathbf{\Lambda}, \mathbf{P}, \mathbf{B}} \frac{1}{|\mathcal{T}|} \sum_{\forall \tau_{t} \in \mathcal{T}} \frac{1}{\left|V_{e}^{t}\right|} \sum_{\forall v_{g} \in V_{e}^{t}} \operatorname{AoV}_{g}\\
		\text { s.t. }
		\operatorname{C1}&: \lambda_{i j}^{t} \in [ \lambda_{i j}^{\min }, \lambda_{i j}^{\max }], \forall D_{i j} \in D_{i}, \forall s_{i} \in S, \forall \tau_{t} \in \mathcal{T}\\
		\operatorname{C2}&: p_{i j}^{t} \in(0,1), \forall D_{i j} \in D_{i}, \forall s_{i} \in S, \forall \tau_{t} \in \mathcal{T}\\
		\operatorname{C3}&: b_{i}^{t} \in [0,b_{e}], \forall s_{i} \in S_{e}^{t}, \forall \tau_{t} \in \mathcal{T}\\
		\operatorname{C4}&: \sum_{\forall D_{i j} \in D_{i}} \lambda_{i j}^{t} \cdot \mathds{E}\left[\operatorname{ser}_{i j}\right]<1, \forall s_{i} \in S, \forall \tau_{t} \in \mathcal{T} \\
		\operatorname{C5}&: \sum_{\forall s_{i} \in S_{e}^{t}} b_{i}^{t} \leq b_{e}, \forall \tau_{t} \in \mathcal{T}
	\end{aligned}
\end{equation}

Constraint $\operatorname{C1}$ requires that the sensing frequencies of information in vehicle $s_{i}$ at time $t$ should meet the requirement of its sensing ability.
$\operatorname{C2}$ guarantees the uploading priority of information in vehicle $s_{i}$ at time $t$.
$\operatorname{C3}$ states that the V2I bandwidth allocated by the edge node $e$ for vehicle $s_{i}$ at time $t$ cannot exceed its bandwidth capacity $b_e$.
$\operatorname{C4}$ guarantees the queue steady-state during the scheduling period $T$. 
$\operatorname{C5}$ requires that the sum of V2I bandwidth allocated by the edge node $e$ cannot exceed its capacity $b_e$.

\section{Proposed Solution}

\begin{figure}
\centering
  \includegraphics[width=0.99\columnwidth]{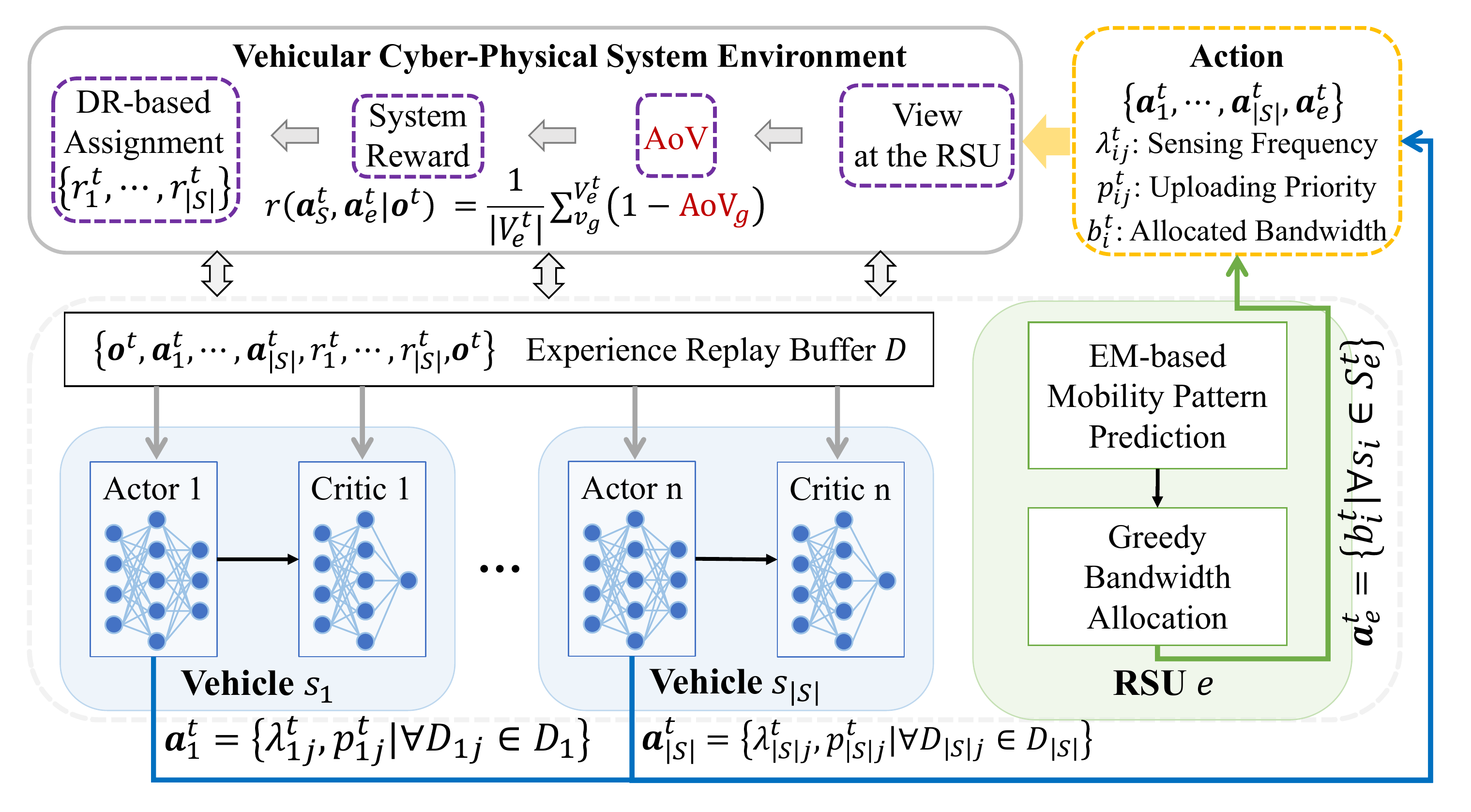}
  \caption{Solution model}
  \label{fig_3_HRL}
\end{figure}

We propose a new solution called Multi-agent Difference-Reward-based deep reinforcement learning with Greedy Bandwidth Allocation (MDR-GBA). 
As shown in Fig. \ref{fig_3_HRL}, the MDR-GBA includes three parts, namely, initialization, replay experiences storing, and training. 
In the initialization part, each vehicle acts as an independent agent and consists of a local actor, a target actor, a local critic, and a target critic network.
The parameters of local actor and critic network of vehicle $s_{i}$ are denoted by $\theta_{i}^{\mu}$ and $\theta_{i}^{Q}$, respectively.
The parameters of target actor and critic network are denoted by $\theta_{i}^{\mu^{\prime}}$ and $\theta_{i}^{Q^{\prime}}$, respectively.
The parameters of each local actor and critic network of vehicles are randomly initialized.
The parameters of target networks are initialized as the same with the corresponding local networks, $\theta_{i}^{\mu^{\prime}} \leftarrow \theta_{i}^{\mu}$, $\theta_{i}^{Q^{\prime}} \leftarrow \theta_{i}^{Q}, \forall s_{i} \in S$.
An experiment replay buffer $D$ with a maximum size $|D|$ is initialized to store replay experiences.

In the replay experiences storing part, the vehicles and RSU decide the actions and store the experiences of interactions between the physical environment.
At the beginning of each iteration, a random process $\mathcal{N}$ is initialized for exploration, and the system status is $\boldsymbol{o}_{1}$.
The RSU broadcasts its view requirements of particular applications and cached information. 
The partial observation of physical environments by vehicle $s_i$ at time  $\tau_t$ is denoted by $\boldsymbol{o}_{i}^{t}=\left\{D_{i}^{t}, D_{e}^{t}, V_{e}^{t}\right\}$, where $D_{i}^{t}$ represents the set of sensed heterogeneous information in vehicle $s_i$ at time $\tau_t$, and $D_{i}^{t} \subseteq D_{i}$.
$D_{e}^{t}$ represents the set of cached information in the RSU, and $D_{e}^{t} \subseteq D$.
$V_{e}^{t}$ represents the set of views required by a particular application at time $\tau_t$.
Hence, the system status at time $\tau_{t}$ is denoted by $\boldsymbol{o}^{t}=\left\{D_{1}^{t}, \cdots, D_{i}^{t}, \cdots, D_{|S|}^{t}, D_{e}^{t}, V_{e}^{t}\right\}$.

The action space of vehicle $s_{i}$ at time $\tau_t$ consists of the sensing frequency and uploading priority of the $j$-th category information sensed by vehicle $s_i$.
The action of vehicle $s_i$ at time $\tau_{t}$ is $\boldsymbol{a}_{i}^{t}=\boldsymbol{\mu}_{\boldsymbol{i}}\left(\boldsymbol{o}_{i}^{t} \mid \theta_{i}^{\mu}\right)+\mathcal{N}_{t}$, where $\mathcal{N}_{t}$ is exploration noise for increasing the diversity of vehicle actions.
The set of actions of vehicles is denoted by $\boldsymbol{a}_{S}^{t} = \left\{\boldsymbol{a}_{i}^{t}\mid \forall s_{i} \in S\right\}$.
Then, the RSU selects the action $\boldsymbol{a}_{e}^{t}=\{b_{i}^{t} \mid \forall s_{i} \in S_{e}^{t}\}$ on determining the V2I bandwidth allocation according to a Greedy Bandwidth Allocation (GBA) scheme.
The set of heterogeneous information sensed by vehicle $s_i$ and required by views $V_{e}^{t}$ at time $\tau_{t}$ is denoted by $D_{i, Req}^{t} = \{ D_{i j} \mid  v_{i j} = 1, \forall v_{i j} \in v_{g}, \forall v_{g} \in V_{e}^{t}, \forall D_{i j} \subseteq D_{i} \}$, and its size can be obtained by $\|D_{i, Req}^{t}\| = \sum_{\forall d_{j k} \in D_{i, Req}^{t}}|d_{j k}|$.
The mobility patterns of vehicles are predicted using Expectation-Maximization (EM) method \cite{hofmann2001unsupervised} with historical relative distances between vehicles and the RSU.
Furthermore, the locations of vehicle $s_i$ at the time interval $[\tau_{t+1}, \tau_{t+h}]$ is predicted based on the EM-based mobility pattens prediction $\operatorname{Traj}_{i}^{t} = \left\{ \hat{l}_{i}^{t+1}, \hat{l}_{i}^{t+2}, \dots, \hat{l}_{i}^{t+h}\right\}$, where $\hat{l}_{i}^{t+1}$ is the predicted location of vehicle $s_i$ at time  $\tau_{t+1}$.
Thus, the average distance of vehicle between RSU is computed by $\operatorname{\bar{dis}}_{i e}^{t} = 1/\left|\operatorname{Traj}_{i}^{t}\right|\cdot\sum_{\forall x \in [1, h]} \widehat{\operatorname{dis}}_{i e}^{t+x}$, where $\widehat{\operatorname{dis}}_{i e}^{t+x}$ is the distance between the predicted location of vehicle $s_i$ and the RSU, $\widehat{\operatorname{dis}}_{i e}^{t+x}=\operatorname{distance}\left(\hat{l}_{i}^{t+x}, l_{e}\right)$.
The allocated bandwidth of vehicle $s_i$ for V2I communications is calculated by $b_{i}^{t} =\frac{b_{e}} {\omega+\operatorname{rank}_{i}}$, where $\omega$ is a constant, and $\operatorname{rank}_{i}$ is the sort ranking by the sequence of $\left\| D_{i,R}^{t}\right\|$ in descending order, and $\operatorname{\bar{dis}}_{i e}^{t}$ in ascending order.

With the determined actions of vehicles and the RSU, a system reward $r\left(\boldsymbol{a}_{S}^{t},\boldsymbol{a}_{e}^{t} \mid \boldsymbol{o}^{t}\right)$ and a new system status $\boldsymbol{o}^{t+1}$ of next time slot are received.
In view of the system optimization objective, which is to minimize the average AoV, we define the average transformed AoV (i.e., $1 -\operatorname{AoV}_{g}$) in a certain time slot as the system reward, which is represented by $r\left(\boldsymbol{a}_{S}^{t},\boldsymbol{a}_{e}^{t} \mid \boldsymbol{o}^{t}\right)=\frac{1}{\left|V_{k}^{t}\right|} \sum_{v_{g}}^{V_{k}^{t}}\left(1 -\operatorname{AoV}_{g} \right)$.
To evaluate the contribution of each vehicle, it is expected to further assign the system reward into individual rewards for independent vehicles to evaluate their contributions.
Accordingly, the Difference reward (DR) \cite{foerster2018counterfactual}, which  is a popular way to perform credit assignments, of vehicle $s_{i}$, denoted by $r_{i}^{t}$, is defined as the difference between the system reward and the reward achieved without its action:
\begin{equation}
\label{equa_dr_sensor_node}
r_{i}^{t}=r\left(\boldsymbol{a}_{S}^{t},\boldsymbol{a}_{e}^{t} \mid \boldsymbol{o}^{t}\right)-r\left(\boldsymbol{a}_{S-i}^{t},\boldsymbol{a}_{e}^{t} \mid \boldsymbol{o}^{t}\right)
\end{equation}
\noindent where $r\left(\boldsymbol{a}_{S-i}^{t},\boldsymbol{a}_{e}^{t} \mid \boldsymbol{o}^{t}\right)$ is the system reward achieved without the contribution of vehicle $s_{i}$, and it can be obtained by setting null action set for $s_i$.
Then, the set of difference rewards of vehicles is represented by $\boldsymbol{r}_{S}^{t}=\{ r_{i}^{t} \mid \forall s_i \in S\}$.
Finally, the experiences are stored in the replay buffer.

In the training part, each vehicle samples a minibatch of $M$ transitions from experience replay buffer $D$ for actor and critic network training. 
One transition of the $M$ minibatch is denoted by $\left(\boldsymbol{o}_{i}^{m}, \boldsymbol{a}_{S}^{m}, \boldsymbol{r}_{S}^{m}, \boldsymbol{o}_{i}^{m+1}\right)$.
The loss function of local critic network of vehicle $s_{i}$ is computed by:
\begin{equation}
\label{equ_loss_sensor}
	\mathcal{L}\left(\theta_{i}^{Q}\right)=\frac{1}{M} \Sigma_{m}\left(y_{m}-Q_{i}\left(\boldsymbol{o}_{i}^{m}, \boldsymbol{a}_{S}^{m} \mid \theta_{i}^{Q}\right)\right)^{2}
\end{equation}
\noindent where $y_{m}=r_{i}^{m}+\gamma Q_{i}^{\prime}\left(\boldsymbol{o}_{i}^{m+1}, \boldsymbol{a}_{S}^{m+1} \mid \theta_{i}^{Q^{\prime}}\right)$ and the action of vehicle $s_{i}$ at time $\tau_{m+1}$ is given by the target actor network based on next observation $\boldsymbol{a}_{i}^{m+1}=\mu_{i}^{\prime}\left(\boldsymbol{o}_{i}^{m+1} \mid \theta_{i}^{\mu^{\prime}}\right)$, and $\gamma$ is the discount rate.
Then, the parameters of local actor network of vehicle $s_{i}$ are updated via policy gradient.
\begin{equation}
\label{equ_gradient_sensor}
	\text{\footnotesize$\nabla_{\theta_{i}^{\mu}} \mathcal{J} \approx \frac{1}{M} \sum_{m} \nabla_{\boldsymbol{a}_{i}^{m}} Q_{i}\left(\boldsymbol{o}_{i}^{m}, \boldsymbol{a}_{S}^{m} \mid \theta_{i}^{Q}\right) \nabla_{\theta_{i}^{\mu}} \mu_{i}\left(\boldsymbol{o}_{i}^{m+1} \mid \theta_{i}^{\mu}\right)$}
\end{equation}
Vehicles softly update the parameters of target networks,
\begin{equation}
	\begin{aligned}
			\theta_{i}^{\mu^{\prime}} &\leftarrow n_{i} \theta_{i}^{\mu}+(1-n_{i}) \theta_{i}^{\mu^{\prime}}, \forall s_{i} \in S\\
			\theta_{i}^{Q^{\prime}} &\leftarrow n_{i} \theta_{i}^{Q}+(1-n_{i}) \theta_{i}^{Q^{\prime}}, \forall s_{i} \in S
	\end{aligned}
\end{equation}
\noindent where $n_{i} \ll 1, \forall s_{i} \in S $.

\section{Performance Evaluation}

\subsection{Settings}
In this section, we implement a simulation model using real-world vehicle trajectories collected from Didi GAIA Initiative \cite{didi}.
In particular, we extract a $3 km \times 3 km$ area of Qingyang district in Chengdu, China from 8:00 am to 8:05 am on 16 Nov. 2016.
The data sizes are uniformly distributed in the range of [100B, 1MB].
The transmission power of each vehicle is set to 1 mW.
The additive white Gaussian noise, the mean channel fading gain, the second moment of the channel fading gain, and the path loss exponent of transmission between vehicles and the RSU are set to -90 dBm, 2, 0.4, and 3, respectively \cite{sadek2009distributed}.
The bandwidth of the RSU is set to 3 MHz.
The noise uncertainly set is uniformly distributed in the range of [0, 3] dB \cite{tandra2008snr}. 
We implement three comparative algorithms as follows.
\begin{itemize}
	\item \textit{Centralized Deep Deterministic Policy Gradient} (C-DDPG) \cite{lillicrap2015continuous}: it implements an agent at the edge node to determine the sensing frequencies, uploading priorities, and V2I bandwidth allocation in a centralized way based on the system state. Meanwhile, the system reward is received by the agent to evaluate its contribution.
	\item \textit{Multi-agent Actor-Critic} (MAC) \cite{lowe2017multi}: it implements agents in vehicles to decide the sensing frequencies and uploading priorities based on local observation of the physical environment, and an agent in the edge node to decide the V2I bandwidth allocation. The system reward is received by each agent to evaluate their contributions, which is the same for each agent.
	\item \textit{MAC with Greedy Bandwidth Allocation} (MAC-GBA): In order to make MAC better at V2I bandwidth allocation, we further design a variation called MAC-GBA, where the RSU allocates the V2I bandwidth based on a greedy scheme by considering vehicle mobility and particular ITS applications requirements.
\end{itemize}

Furthermore, the following metrics are designed for performance evaluation.
\begin{itemize}
	\item \textit{Cumulative Reward} (CR): it is the cumulative system reward in the scheduling period $\mathcal{T}$, which is computed by $ \sum_{\forall \tau_{t} \in \mathcal{T}} r\left(\boldsymbol{a}_{S}^{t},\boldsymbol{a}_{e}^{t} \mid \boldsymbol{o}^{t}\right)$.
	\item \textit{Composition of Average Reward} (CAR): it is defined as the percentage of the normalized timeliness, completeness, and consistency in the average reward and formulated by $<\frac{3}{10} (1-\hat{\theta_{g}}),\frac{4}{10}\chi_{g},\frac{3}{10}(1-\hat{\xi_{g}})>$.
	\item \textit{Average Queuing Time} (AQT): it is defined as the sum of queuing time of the sensed information divided by the number of information, which is computed by $ \sum_{\forall \tau_{t} \in \mathcal{T}}\{\{ \sum_{s_{i} \in S} \{\sum_{\forall D_{i j} \subseteq D_{i}}{wai}_{i j}\} / |D_{i}| \} / |S| \} / |\mathcal{T}|$.
	\item \textit{Service Ratio} (SR): it is defined as the number of views which satisfy the completeness requirement over the total number of required views during the scheduling period $\mathcal{T}$, and it is computed by $\sum_{\forall \tau_{t} \in \mathcal{T}}\sum_{\forall v_{g} \in V_{e}^{t}} \mathds{1}\{\chi_{g} \geq \chi_{threshold}\}  / \sum_{\forall \tau_{t} \in \mathcal{T}} |V_{e}^{t}|$.
\end{itemize}

\subsection{Results and Analysis}

\begin{figure*}[t!]
  \centering
  \subfloat[CR]{\includegraphics[height=1.22in]{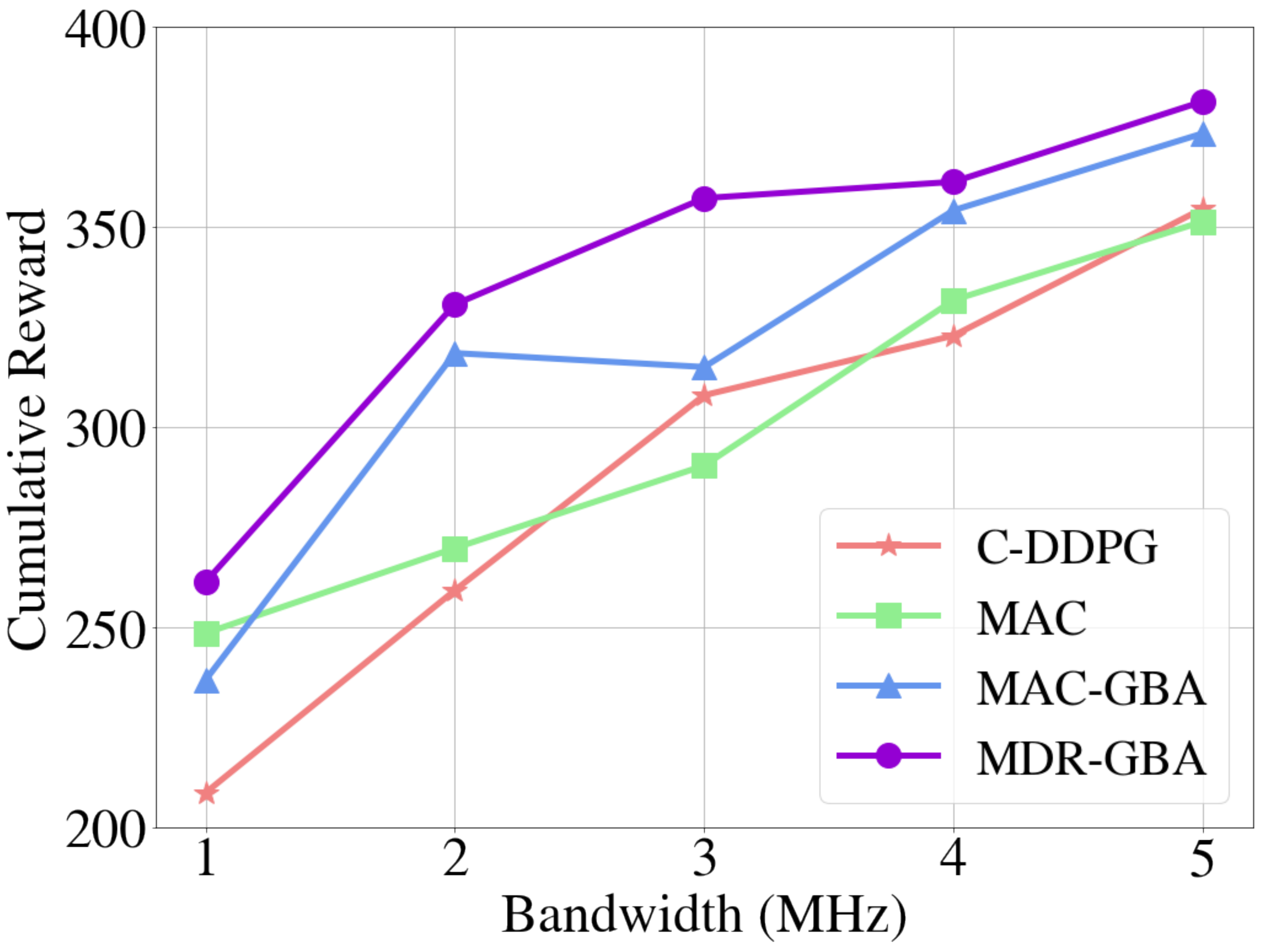}}
  \hspace{0.5em}
  \subfloat[CAR]{\includegraphics[height=1.22in]{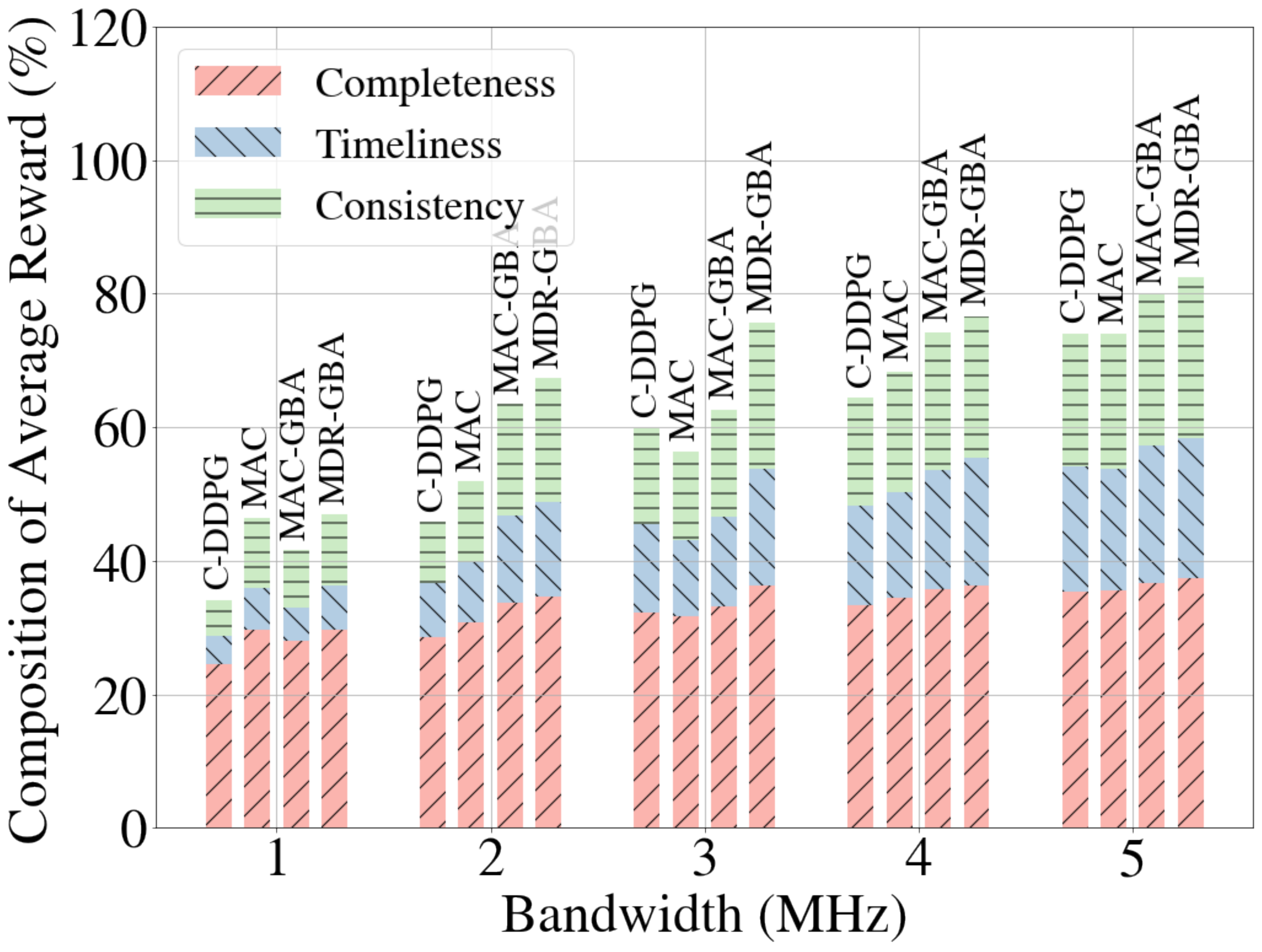}}
  \hspace{0.5em}
  \subfloat[AQT]{\includegraphics[height=1.22in]{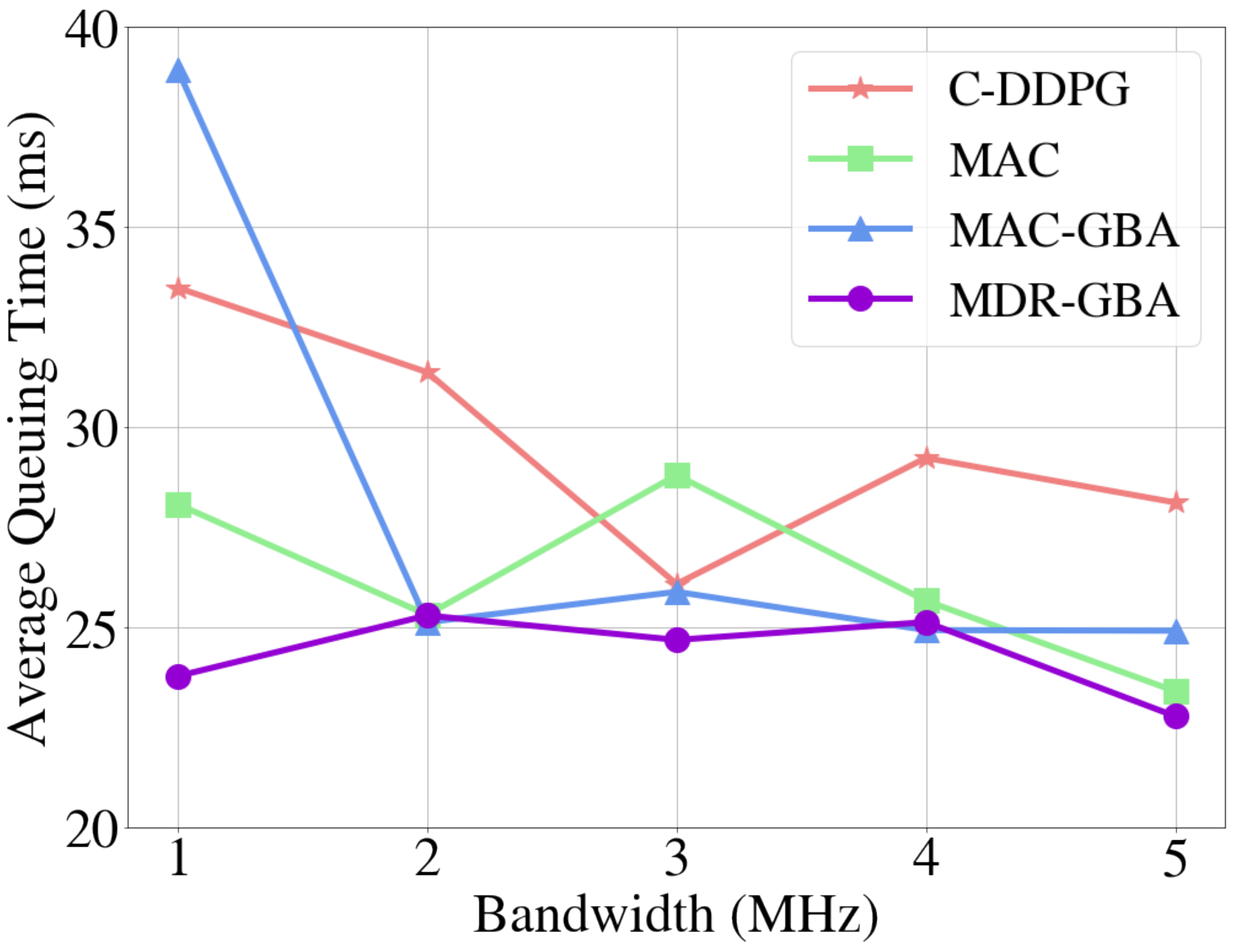}}
  \hspace{0.5em}
  \subfloat[SR]{\includegraphics[height=1.22in]{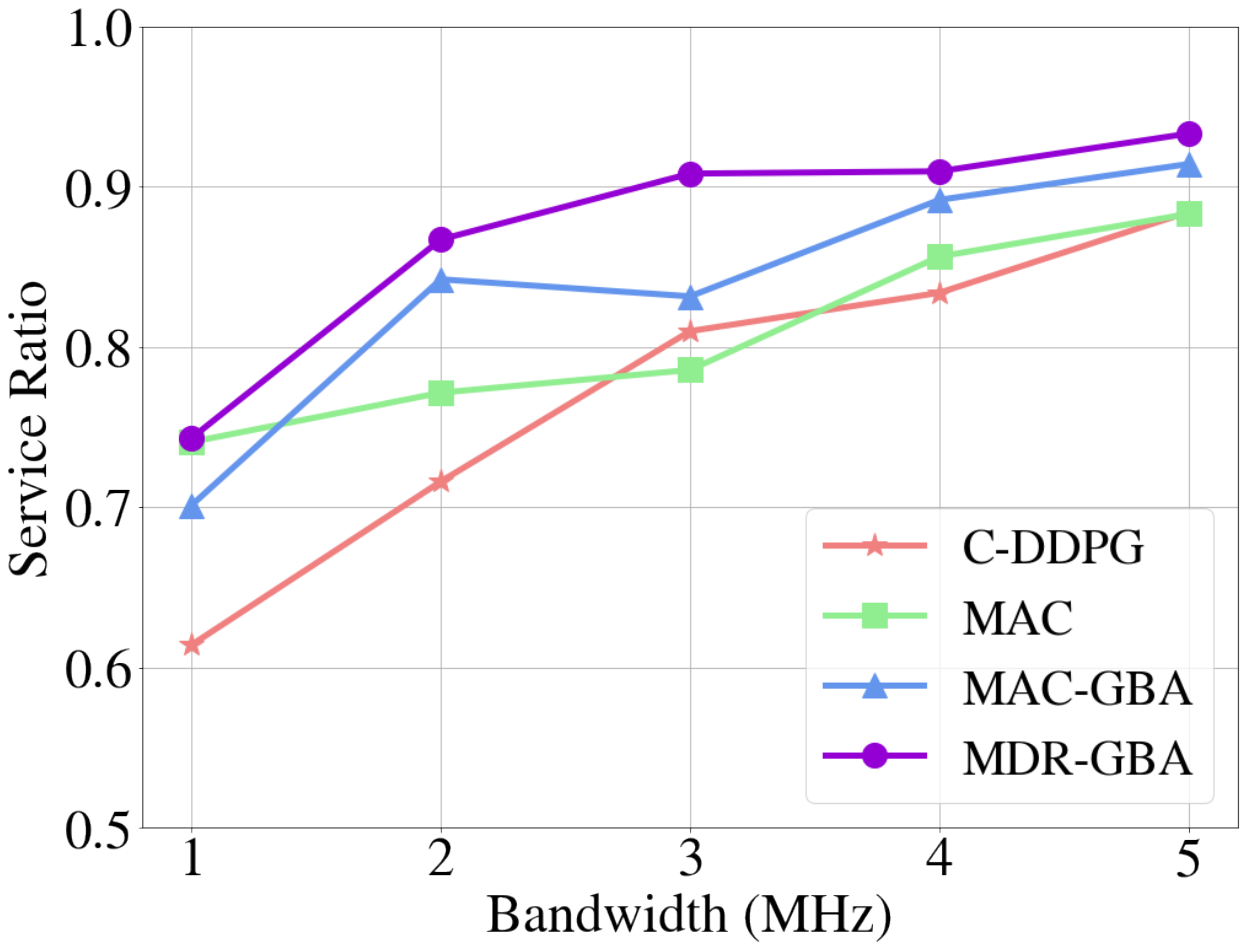}}
  \caption{Performance comparison under different RSU bandwidths}
  \label{fig_5_bandwidth}
\end{figure*}

\begin{figure*}[t!]
  \centering
  \subfloat[CR]{\includegraphics[height=1.22in]{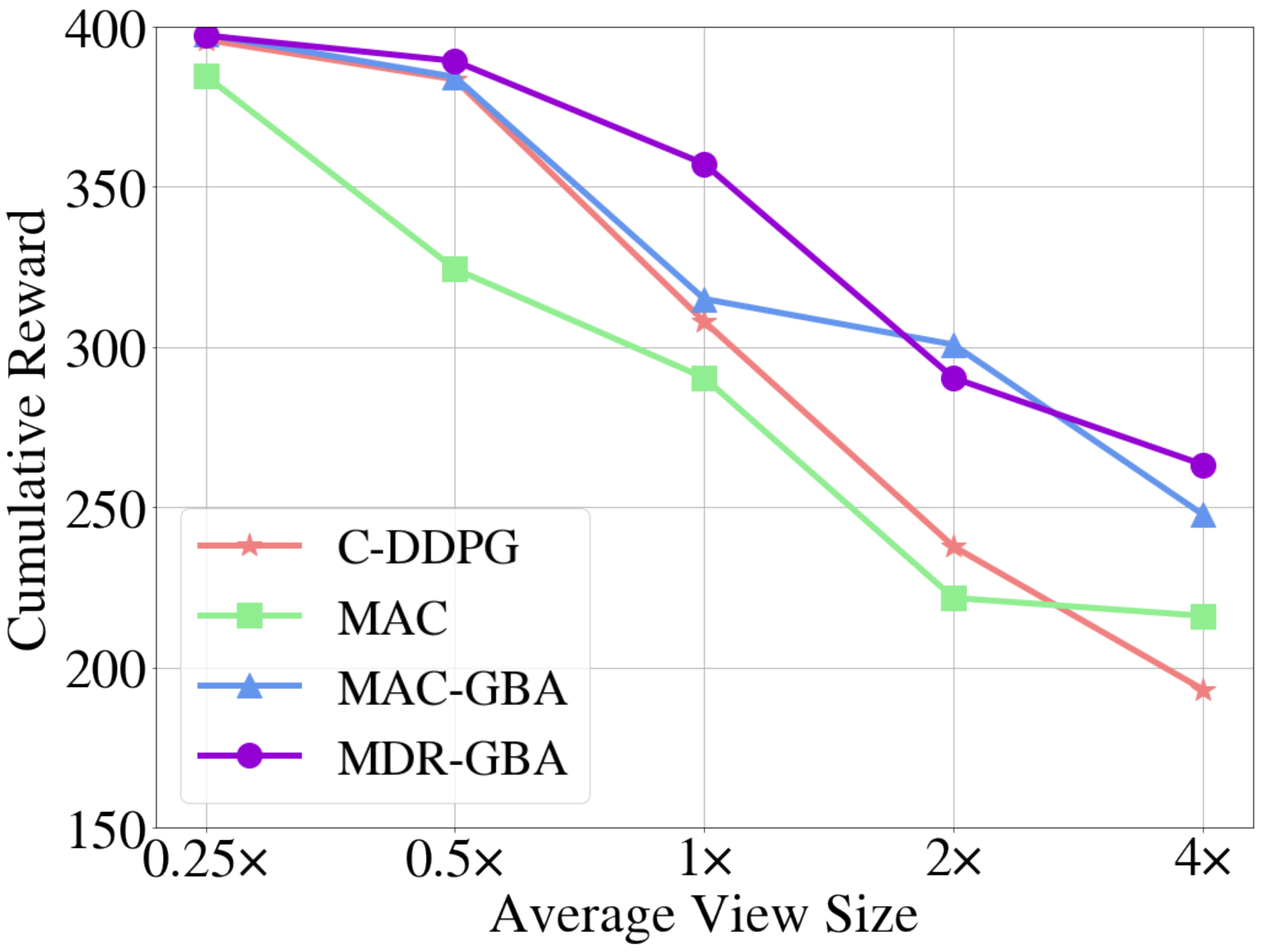}}
  \hspace{0.5em}
  \subfloat[CAR]{\includegraphics[height=1.22in]{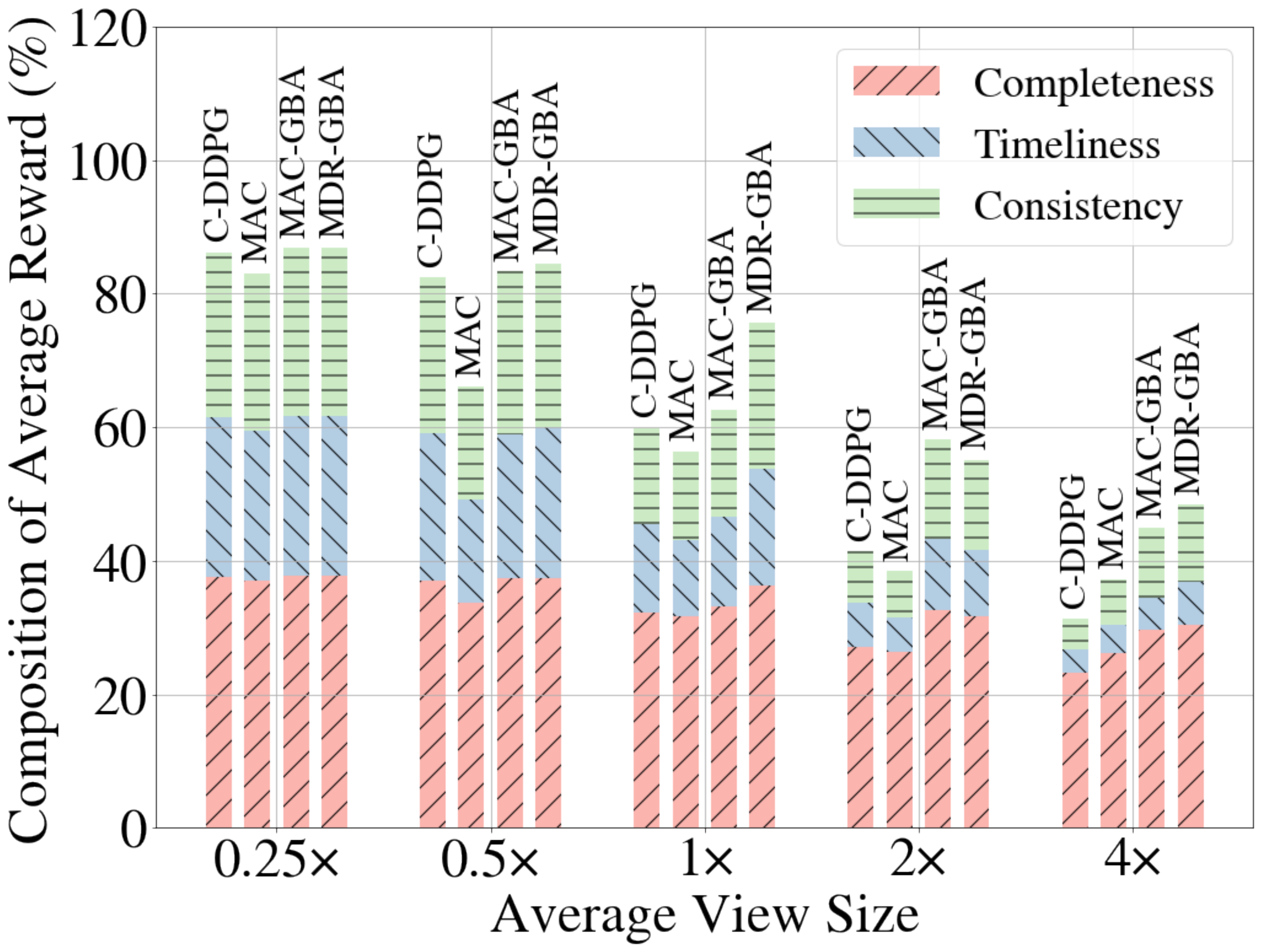}}
  \hspace{0.5em}
  \subfloat[AQT]{\includegraphics[height=1.22in]{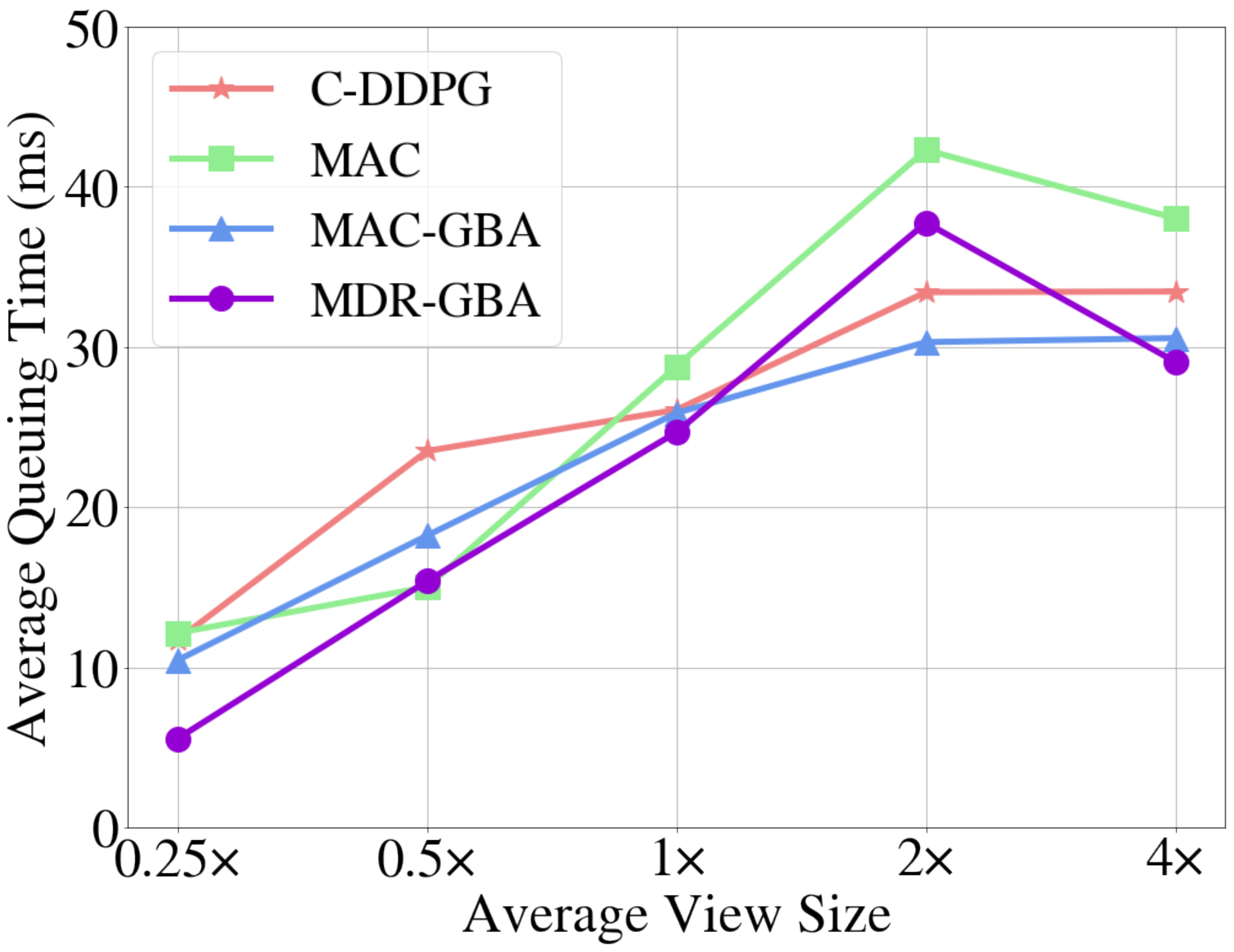}}
  \hspace{0.5em}
  \subfloat[SR]{\includegraphics[height=1.22in]{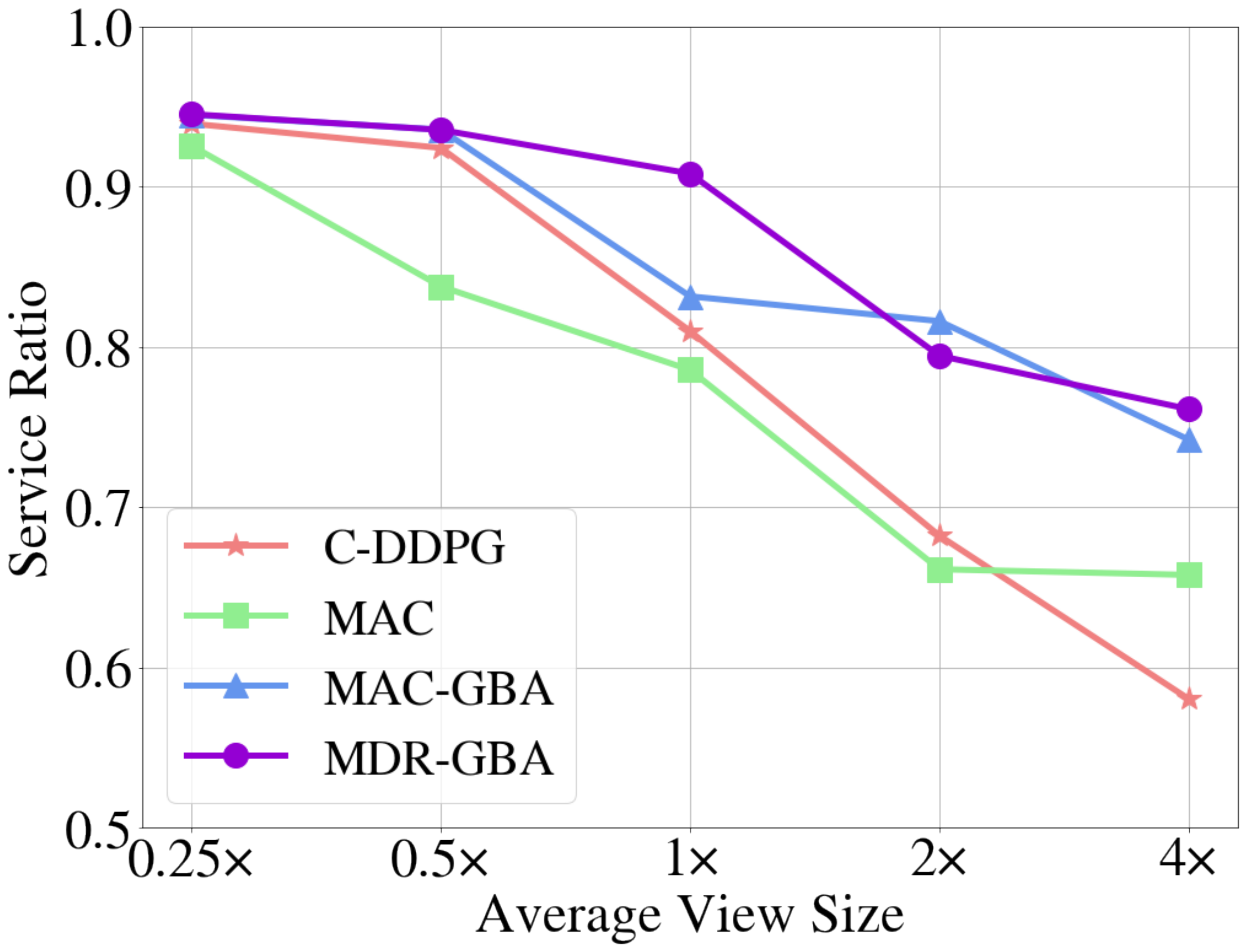}}
  \caption{Performance comparison under different application requirements on views}
  \label{fig_6_datasize}
\end{figure*}

\textit{1) Effect of RSU bandwidth:}
Fig. 4(a) compares the CR of algorithms under different RSU bandwidths.
The CR of MDR-GBA outperforms C-DDPG, MAC, and MAC-GBA at around 17.63\%, 12.62\%, and 6.34\%, respectively.
It is observed that the difference between MDR-GBA and MAC-GBA decreases when the bandwidth increases from 4 MHz to 5 MHz. 
The reason is that when there is sufficient bandwidth for data uploading, the scheduling significance is getting weaker.
Fig. 4(b) compares the CAR, and MDR-GBA achieves better performance than the other three algorithms.
This is mainly because the cooperation of sensing and uploading information among vehicles is more efficient in MDR-GBA under the limited bandwidth.
Fig. 4(c) compares the AQT, as noted, the AQT of MDR-GBA maintains the lowest under different RSU bandwidths, which reflects that the designed GBA scheme can allocate the bandwidth more efficiently. 
The advantage can be further justified by Fig. 4(d), which shows the SR of the algorithms.
The SR of MDR-GBA remains at the highest level across all the cases.

\textit{2) Effect of application requirements on views:} Fig. 5(a) compares the CR of the algorithms under different application requirements on views.
The CR of MDR-GBA outperforms C-DDPG, MAC, and MAC-GBA at around 15.25\%, 19.80\%, and 3.49\%, respectively.
Note that the CR of MDR-GBA, MAC-GBA, and MAC is similar when the average view size is small (i.e. around 1.62 MB). 
It is because a smaller data size has a higher probability to be successfully uploaded. 
Fig. 5(b) compares the CAR, and it is observed that the performance difference between MDR-GBA and MAC-GBA is small when the average view size increases from 0.25$\times$ to 0.5$\times$.
This is because the scheduling effect is insignificant when sufficient resources meet the requirements of a smaller average view size (i.e., around 1.62 MB and 3.23 MB).
Figs. 5(c) and 5(d) compare the AQT and SR, showing that the MDR-GBA can remain the lowest AQT, and meanwhile achieve the highest SR in most cases.
It is noted that MAC-GBA achieves the lowest AQT and the highest SR when the average view size is 2$\times$, which reflects that the GBA scheme can allocate the bandwidth more efficiently. 

\section{Conclusion}
In this work, we designed a new metric AoV to evaluate the quality of heterogeneous information fusion in VCPS.
In particular, a multi-class M/G/1 priority queue was adopted to model the queuing of sensed information at vehicles, and the Shannon theory was applied to model the V2I-based information uploading.
A solution called MDR-GBA was proposed, in which vehicles act as independent agents and decide both sensing frequencies and uploading priorities.
A greedy scheme was designed for V2I bandwidth allocation at the RSU. 
Finally, we build the simulation model, and the results demonstrate the significance of the AoV and the superiority of MDR-GBA.

%\section*{Acknowledgment}
%This work was supported in part by the National Natural Science Foundation of China under Grant Nos. 62172064 and 62171072, the Chongqing Young-Talent Program (Project No. cstc2022ycjh-bgzxm0039), and the Venture \& Innovation Support Program for Chongqing Overseas Returnees (Project No. cx2021063).

\bibliographystyle{IEEEtran} 
\bibliography{reference}

% Generated by IEEEtran.bst, version: 1.12 (2007/01/11)
\begin{thebibliography}{10}
\providecommand{\url}[1]{#1}
\csname url@samestyle\endcsname
\providecommand{\newblock}{\relax}
\providecommand{\bibinfo}[2]{#2}
\providecommand{\BIBentrySTDinterwordspacing}{\spaceskip=0pt\relax}
\providecommand{\BIBentryALTinterwordstretchfactor}{4}
\providecommand{\BIBentryALTinterwordspacing}{\spaceskip=\fontdimen2\font plus
\BIBentryALTinterwordstretchfactor\fontdimen3\font minus
  \fontdimen4\font\relax}
\providecommand{\BIBforeignlanguage}[2]{{%
\expandafter\ifx\csname l@#1\endcsname\relax
\typeout{** WARNING: IEEEtran.bst: No hyphenation pattern has been}%
\typeout{** loaded for the language `#1'. Using the pattern for}%
\typeout{** the default language instead.}%
\else
\language=\csname l@#1\endcsname
\fi
#2}}
\providecommand{\BIBdecl}{\relax}
\BIBdecl

\bibitem{jia2015survey}
D.~Jia, K.~Lu, J.~Wang, X.~Zhang, and X.~Shen, ``A survey on platoon-based
  vehicular cyber-physical systems,'' \emph{IEEE Commun. Surv. Tutor.},
  vol.~18, no.~1, pp. 263--284, 2015.

\bibitem{liu2020fog}
K.~Liu, K.~Xiao, P.~Dai, V.~C. Lee, S.~Guo, and J.~Cao, ``Fog computing
  empowered data dissemination in software defined heterogeneous vanets,''
  \emph{IEEE. Trans. Mob. Comput.}, vol.~20, no.~11, pp. 3181--3193, 2021.

\bibitem{singh2020intent}
A.~Singh, G.~S. Aujla, and R.~S. Bali, ``Intent-based network for data
  dissemination in software-defined vehicular edge computing,'' \emph{IEEE
  Trans. Intell. Transp. Syst.}, vol.~22, no.~8, pp. 5310--5318, 2020.

\bibitem{dai2020deep}
Y.~Dai, D.~Xu, K.~Zhang, S.~Maharjan, and Y.~Zhang, ``Deep reinforcement
  learning and permissioned blockchain for content caching in vehicular edge
  computing and networks,'' \emph{IEEE Trans. Veh. Technol.}, vol.~69, no.~4,
  pp. 4312--4324, 2020.

\bibitem{xiao2021cooperative}
K.~Xiao, K.~Liu, X.~Xu, L.~Feng, Z.~Wu, and Q.~Zhao, ``Cooperative coding and
  caching scheduling via binary particle swarm optimization in software-defined
  vehicular networks,'' \emph{Neural. Comput. Appl.}, vol.~33, no.~5, pp.
  1467--1478, 2021.

\bibitem{shang2021deep}
B.~Shang, L.~Liu, and Z.~Tian, ``Deep learning-assisted energy-efficient task
  offloading in vehicular edge computing systems,'' \emph{IEEE Trans. Veh.
  Technol.}, vol.~70, no.~9, pp. 9619--9624, 2021.

\bibitem{liao2020learning}
H.~Liao, Z.~Zhou, W.~Kong, Y.~Chen, X.~Wang, Z.~Wang, and S.~Al~Otaibi,
  ``Learning-based intent-aware task offloading for air-ground integrated
  vehicular edge computing,'' \emph{IEEE Trans. Intell. Transp. Syst.},
  vol.~22, no.~8, pp. 5127--5139, 2021.

\bibitem{zhang2019cyber}
Y.~Zhang, L.~Chu, Y.~Ou, C.~Guo, Y.~Liu, and X.~Tang, ``A cyber-physical
  system-based velocity-profile prediction method and case study of application
  in plug-in hybrid electric vehicle,'' \emph{IEEE T. Cybern.}, vol.~51, no.~1,
  pp. 40--51, 2019.

\bibitem{zhang2020data}
T.~Zhang, Y.~Zou, X.~Zhang, N.~Guo, and W.~Wang, ``Data-driven based cruise
  control of connected and automated vehicles under cyber-physical system
  framework,'' \emph{IEEE Trans. Intell. Transp. Syst.}, vol.~22, no.~10, pp.
  6307--6319, 2020.

\bibitem{lian2020cyber}
Y.~Lian, Q.~Yang, W.~Xie, and L.~Zhang, ``Cyber-physical system-based heuristic
  planning and scheduling method for multiple automatic guided vehicles in
  logistics systems,'' \emph{IEEE Trans. Ind. Inform.}, vol.~17, no.~11, pp.
  7882--7893, 2021.

\bibitem{liu2014temporal}
K.~Liu, V.~C.~S. Lee, J.~K.-Y. Ng, J.~Chen, and S.~H. Son, ``Temporal data
  dissemination in vehicular cyber--physical systems,'' \emph{IEEE Trans.
  Intell. Transp. Syst.}, vol.~15, no.~6, pp. 2419--2431, 2014.

\bibitem{liu2014scheduling}
K.~Liu, V.~C. Lee, J.~K. Ng, S.~H. Son, and E.~H.-M. Sha, ``Scheduling temporal
  data with dynamic snapshot consistency requirement in vehicular
  cyber-physical systems,'' \emph{ACM Trans. Embed. Comput. Syst.}, vol.~13,
  no.~5s, pp. 1--21, 2014.

\bibitem{xu2020vehicular}
X.~Xu, K.~Liu, K.~Xiao, L.~Feng, Z.~Wu, and S.~Guo, ``Vehicular fog computing
  enabled real-time collision warning via trajectory calibration,''
  \emph{Mobile Netw. Appl.}, vol.~25, no.~6, pp. 2482--2494, 2020.

\bibitem{lv2018driving}
C.~Lv, X.~Hu, A.~Sangiovanni-Vincentelli, Y.~Li, C.~M. Martinez, and D.~Cao,
  ``Driving-style-based codesign optimization of an automated electric vehicle:
  A cyber-physical system approach,'' \emph{IEEE Trans. Ind. Electron.},
  vol.~66, no.~4, pp. 2965--2975, 2018.

\bibitem{lillicrap2015continuous}
T.~P. Lillicrap, J.~J. Hunt, A.~Pritzel, N.~Heess, T.~Erez, Y.~Tassa,
  D.~Silver, and D.~Wierstra, ``Continuous control with deep reinforcement
  learning,'' \emph{arXiv preprint, arXiv:1509.02971}, 2015.

\bibitem{lowe2017multi}
R.~Lowe, Y.~I. Wu, A.~Tamar, J.~Harb, O.~Pieter~Abbeel, and I.~Mordatch,
  ``Multi-agent actor-critic for mixed cooperative-competitive environments,''
  in \emph{Proc. of Neural Information Process. Syst. (NeurIPS)}, vol.~30,
  2017.

\bibitem{didi}
``Data source: Didi chuxing gaia open dataset initiative,''
  \url{https://outreach.didichuxing.com/research/opendata/en/}.

\bibitem{moltafet2020age}
M.~Moltafet, M.~Leinonen, and M.~Codreanu, ``On the age of information in
  multi-source queueing models,'' \emph{IEEE Trans. Commun.}, vol.~68, no.~8,
  pp. 5003--5017, 2020.

\bibitem{sadek2009distributed}
A.~K. Sadek, Z.~Han, and K.~R. Liu, ``Distributed relay-assignment protocols
  for coverage expansion in cooperative wireless networks,'' \emph{IEEE. Trans.
  Mob. Comput.}, vol.~9, no.~4, pp. 505--515, 2009.

\bibitem{tandra2008snr}
R.~Tandra and A.~Sahai, ``Snr walls for signal detection,'' \emph{IEEE J. Sel.
  Top. Signal Process.}, vol.~2, no.~1, pp. 4--17, 2008.

\bibitem{hofmann2001unsupervised}
T.~Hofmann, ``Unsupervised learning by probabilistic latent semantic
  analysis,'' \emph{Mach. Learn.}, vol.~42, no.~1, pp. 177--196, 2001.

\bibitem{foerster2018counterfactual}
J.~N. Foerster, G.~Farquhar, T.~Afouras, N.~Nardelli, and S.~Whiteson,
  ``Counterfactual multi-agent policy gradients,'' in \emph{Proc. AAAI Conf. on
  Artif. Intell. (AAAI)}, 2018.

\end{thebibliography}

\end{document}